\documentclass[11pt,twoside]{article}
\usepackage{asp2010}
\resetcounters

\begin{document}


\title{Dark Matter and its Effects on Helioseismology}
\author{Ryan Hamerly and Alexander Kosovichev
\affil{Stanford University, Stanford, CA 94305, USA}}

\begin{abstract}
Helioseismology can be used to place new constraints on the properties of dark matter, allowing solar observations to complement more conventional dark matter searches currently in operation.  During the course of its lifetime, the Sun accretes a sizeable amount of dark matter.  This accreted matter affects the heat transport of the stellar core in ways that helioseismology can detect, or at least constrain.  We modify the CESAM stellar evolution code to take account of dark matter and determine the effect of WIMP models on the stellar structure and normal-mode oscillation frequencies.
\end{abstract}

\section{Introduction}
It has been known since the 1930's that the galaxy is permeated with a halo of ``dark matter'' which thus far has only been detected gravitationally.  Many theories of physics, including supersymmetric extensions to the standard model, postulate the existence of stable Weakly Interacting Massive Particles (WIMP's).  Most WIMP's have masses in the GeV range, interact through the weak and gravitational forces, and make promising dark matter candidates.
As dark matter particles in the Sun's vicinity pass through the Sun, a fraction of them will scatter off of nuclei in the Sun and become captured by the Sun's gravitational field.  These particles quickly thermalize to form a dark matter ``halo'' in the solar interior.  The size of this halo depends on the WIMP's mass and the temperature and density at the center of the Sun:

\[
r_\chi   = \sqrt {\frac{{3kT_c }}{{2\pi G\rho _c m_\chi  }}}
\]

Typically, the size of this halo ranges from 0.01 -- 0.05 solar radii, so the dark matter halo is buried deep inside the Sun.
The number of particles $N_\chi$ inside the Sun's dark matter halo will grow due to the above mentioned capture process, and shrink due to self-annihilation of dark matter particles. $N_\chi$ thus evolves as:

\[
\dot N_\chi   = A - B{\kern 1pt} N_\chi ^2,
\]

where

\[A = 3.14 \times 10^{39} \left( {\frac{{\rho _\chi  }}{{\rm GeV/cm^3 }}} \right)\left( {\frac{{m_\chi  }}{{\rm GeV}}} \right)^{ - 1} \frac{1}{{\rm Myr}}\]
 \[  B = 4.57 \times 10^{ - 13} \left( {\frac{{T_c }}{{\rm 15MK}}} \right)^{1/2} \left( {\frac{{m_\chi  }}{{\rm GeV}}} \right)^{ - 1/2} \left( {\frac{{r_\chi  }}{{r_{\sun} }}} \right)^{-3} \left( {\frac{{\sigma _{\rm ann} }}{{\rm cm^3 }}} \right)\frac{1}{{\rm Myr}} .
\]

This dark matter halo will affect the Sun's interior structure in two ways: first, by creating an additional heat source (self-annihilation), and second, by conducting altering the Sun's heat conductivity.  The former can be neglected by a simple order-of-magnitude analysis, but the latter cannot.  While WIMP-nucleon interactions are very rare in the Sun, the mean free path of the WIMP's is correspondingly much larger than that of the photon, and these two effects can cancel out to give a dark-matter-mediated thermal conductance that equals or exceeds that of ordinary radiative transport -- inside the dark matter halo, that is.

These dark matter effects result in changes in the solar structure, which are subtle but in principle can be detected by helioseismology \citep{1985ApJ...299..994F} and solar neutrino experiments \citep{2010PhRvL.105a1301F}. Recently, \cite{2010ApJ...722L..95L} suggested to search for dark matter effects in frequency shift and spacing of g-modes (which, however, have not been detected).  \cite{2010PhRvD..82j3503C} calculated these effects for low-degree p-mode frequencies, and concluded that these are insignificant. However, while dark matter particles may be concentrated in a very small central region of the Sun their presence would affect the whole structure of the Sun and change not only low-degree oscillation p-modes, but also medium-degree modes, thus, producing characteristic patterns of the frequency shift in the p-mode spectrum. Our ultimate goal is to investigate these patterns for various WIMP models, and develop a procedure for determining constraint on these models from helioseismology observations.

\subsection{Method}
To simulate the dynamics of the Sun, we used CESAM, a stellar evolution code developed by P. Morel and collaborators \citep{1997A&AS..124..597M}.  CESAM was written to study mid-size main-sequence stars like the Sun, and can generate very accurate, calibrated solar models in less than an hour on a typical workstation.  Original versions of the code were written in Fortran 77, but the most recent installment, CESAM2k, is written in F95.
CESAM rigorously segregates the ``numerical space'' -- where the differential equations are solved -- from the ``physical space'' where physical parameters and effects are declared.  This makes the code particularly well adapted to new physics, since the ``physical'' functions can be straightforwardly modified without any knowledge of the numerical details of the simulation.
To account for the thermal transport due to the dark matter, we introduce a modification to the opacity routine in CESAM:

\[
\frac{1}{\kappa } = \frac{1}{{\kappa _{rad} }} + \frac{{f(K_n )}}{{\kappa _{dm} }},
\]
where $K_n = l_\chi/r_\chi$ is the ratio of the mean free path to the scale height, $f(x) = 1/(1 + 6.2x^2)$, and $\kappa_{dm}$ (in g/cm$^2$) is given by:

\[
\frac{1}{{\kappa _{dm}}} = 1.03\times 10^{-46}\left({\frac{{l_\chi  }}{{r_{\sun} }}}\right)\left({\frac{{r_\chi  }}{{r_{\sun}}}} \right)^{-3} \left({\frac{\rho }{{{\rm 150\frac{g}{cm^3} }}}}\right)\left( {\frac{T}{{\rm 15MK}}} \right)^{-5/2} \left( {\frac{{m_\chi  }}{{GeV}}} \right)^{-1/2} N_\chi  \lambda.
\]

Ultimately, we would like the simulations to depend only on the parameters of the dark matter itself -- the galactic WIMP density $\rho_\chi$ and the WIMP mass etc.  Presently, however, we have only calculated stellar models with ad hoc changes to the opacity of the form:

\[
\frac{1}{\kappa } = \frac{1}{{\kappa _{rad} }} + \frac{1}{{\kappa _0 }}e^{ - r^2 /r_\chi ^2 }
\]

We computed eight solar models (Table~1).  In the first four (Models A--D), $\kappa_0$ was held fixed and $r_\chi$ was varied; in the others (Models E--H), it was $\kappa_0$ that was varied.
\begin{table}
\begin{center}
    \begin{tabular}{| c  | c c || c  | c c |}
    \hline
    Model & $\kappa_0$ (cm$^2$/g) & $r_\chi(r_\odot)$ & Model & $\kappa_0$ (cm$^2$/g) & $r_\chi(r_\odot)$ \\
    \hline

    A & 0.00 & 0.00 & E & 0.50 & 0.05 \\
    B & 1.00 & 0.02 & F & 0.75 & 0.05 \\
    C & 1.00 & 0.04 & G & 1.00 & 0.05 \\
    D & 1.00 & 0.06 & H & 2.00 & 0.05\\
    \hline
\end{tabular}
\caption{Models with Varying Scale Height (B--D) and Varying Central DM Opacity (E--H). Model A is the standard solar model without dark matter.}
\end{center}
\label{table1}
\end{table}
The models were calibrated to solar  radius, luminosity, mass, and composition.  Test simulations were run to ensure that the calibration was sufficiently accurate.
\subsection{Results}
\begin{figure}
\begin{center}
\includegraphics[width=\textwidth] {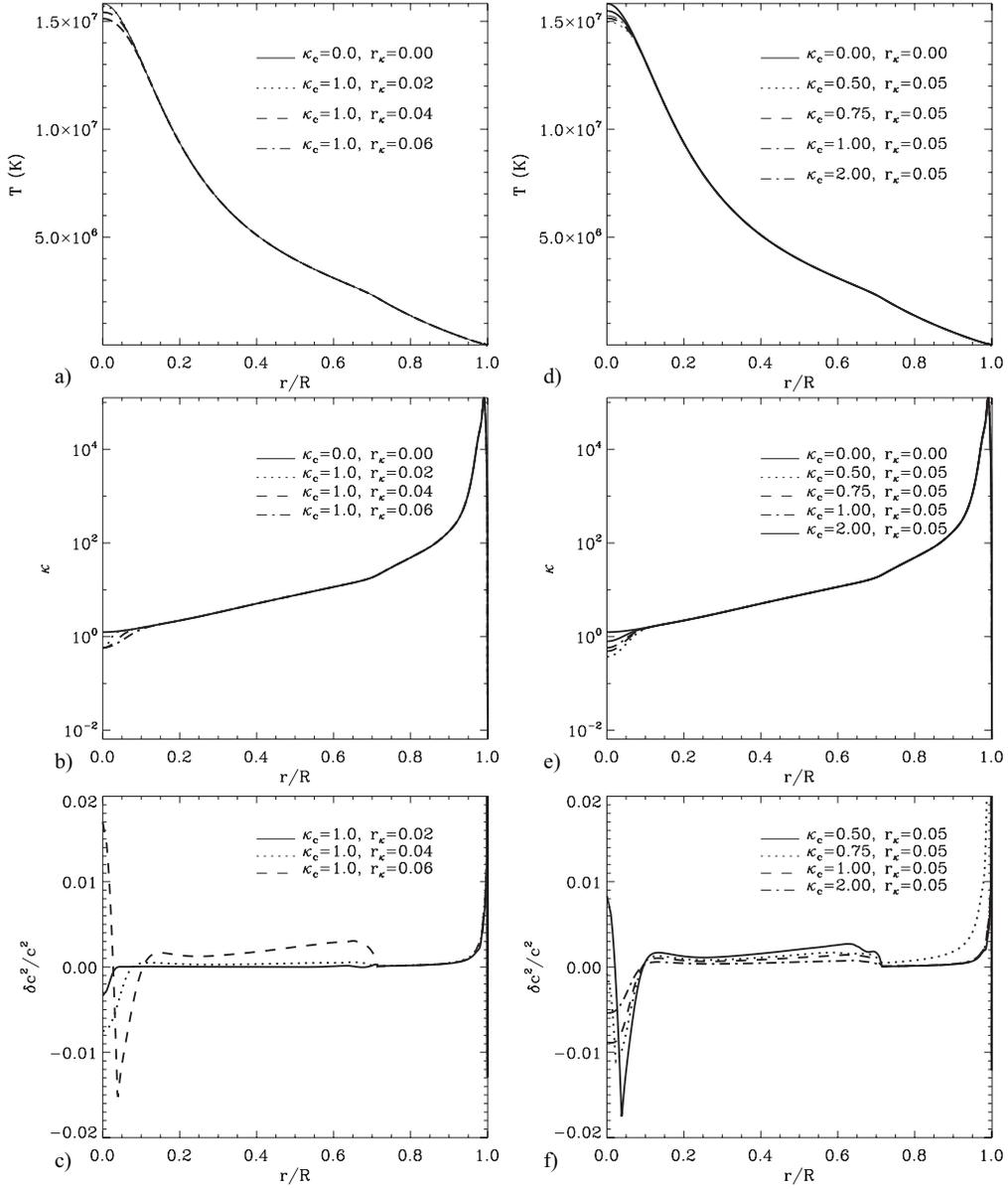}
\end{center}
\caption{Temperature (top), opacity (middle), and sound-speed variations relative to the solar model without dark matter (bottom) for the models with varying dark matter scale-height (Models A--D in Table 1; panels a--c), and for the models with varying dark matter opacity (Models E--H; panels d--f).
\label{fig1}}
\end{figure}

The simulation results for the temperature, opacity and relative sound-speed variations for Models B--H are shown in Figure~\ref{fig1}.  They show the decrease in temperature and opacity in the central part of the Sun where dark matter is concentrated according to our models. The squared sound-speed variations relative to the standard solar model without dark matter also show a decrease in the central core. However, when the dark matter opacity effects are stronger due to higher conductivity $\kappa_0$ or larger $r_\chi$ the sound-speed variations are non-monotonic, and qualitatively resemble the sound-speed profile obtained by helioseismology inversions \citep{1999JCoAM.109....1K}. With improved precision of p-mode frequency measurements these inversions may be able to better resolve the sound-speed structure of the central core, and provide constraints on dark matter properties.

\begin{figure}
\begin{center}
\includegraphics[width=\textwidth] {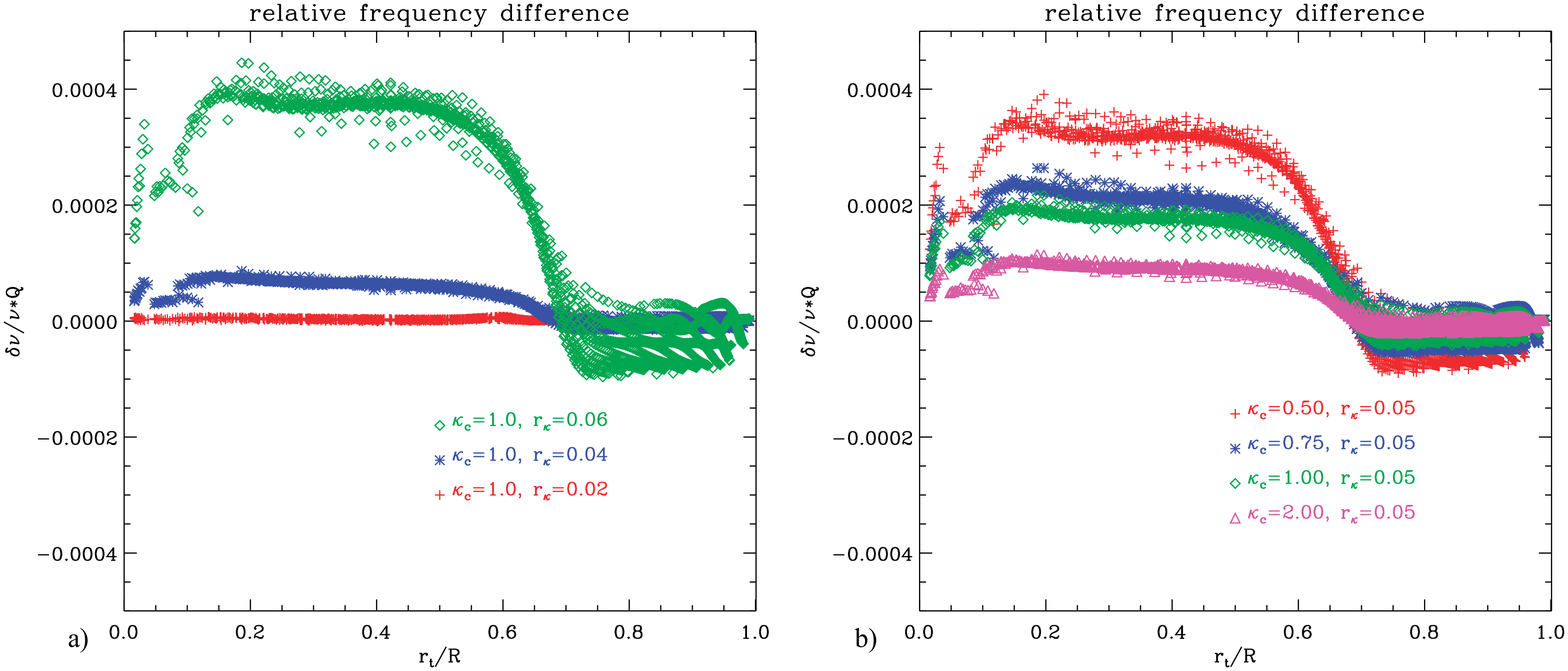}
\end{center}
\caption{Relative frequency differences, scaled with mode inertia $Q$, between the dark matter solar models and the standard solar model as a function of radius $r_t$ of the inner turning point for: a) models with varying scale height (Models B-D), and b) the models with varying dark matter opacity (Models E--H).
\label{fig2}}
\end{figure}

In Figure~\ref{fig2} we plot the relative frequency difference, scaled with mode inertia $Q$, between the models with dark matter and the standard model without dark matter. The scaled frequency is plotted as a function of  the radius of the inner turning point of p-modes, $r_t$, in the observed frequency and angular degree range. In the asymptotic theory, the scaled frequency difference is a sum of a function of  $r_t$ and a function of frequency representing non-adiabatic and atmospheric effects \citep{Kosovichev2011}. In our simulations we calculated the oscillation frequencies in the adiabatic approximation, and assumed that the atmospheric structure is the same for all models. Therefore, the frequency differences in Fig.~\ref{fig2} essentially depend only on the radius of the inner turning point. This dependence shows sharp non-monotonic variations at small values of the radius. In terms of the angular degree this variation represents a transition from low-degree to medium-degree modes. This transition may be an important characteristic for the detection of dark matter effects on the Sun. The medium-degree modes show a systematic shift, which is almost constant for  p-modes with the inner turning points located inside the radiative zone. However, the modes of higher angular degree with the turning points in the convection zone are not affected. These frequency patterns can compared directly with the observed frequencies. The current precision of the frequency measurements is about $10^{-5}$. Therefore, our simulations show that helioseismology may be capable of providing constraints for dark matter models.

\subsection{Conclusions}

These simulations show that opacity modifications similar to those induced by dark matter WIMP's can have profound effects on the heat transport in the inner core of the Sun.  The additional heat transport flattens out the temperature profile in the core, raising its density.  The sound speed in these models changed by as much as a percent, suggesting that WIMP transport may have detectable effects on helioseismic measurements.
To test these claims, it will be necessary to replace our ad hoc opacity prescription with a more physical model which computes $\kappa_0$ and $r_\chi$ as functions of the total number $N_\chi$ of WIMP's in the halo, which evolves as the star ages, and the star's central temperature and density.
Moreover, the flattening of the temperature profile in the core may noticeably change the flux of high-energy solar neutrinos.  Together with helioseismic observations, it is possible that solar neutrino measurements may be used to constrain the WIMP parameter space.


\end{document}